\documentclass[12pt,preprint]{aastex}
\begin{document}

%
\def\simlt {\gtrsim}
\def \etal {{\it et al. }}
\def \bx {{\bf x}}
\def \Mpc {h_{0.65}^{-1}{\rm Mpc}}
\def \kpc {h_{0.65}^{-1}{\rm kpc}}
\def \farcs{\hbox{$.\!\!^{\prime\prime}$}}
\def \farcm{\hbox{$.\!\!^{\prime}$}}
 

\title{Joint X-Ray and Optical Measurements of the Mass 
Distribution of the Distant Galaxy Cluster CLJ~0152.7$-$1357}

\author{Zhi-Ying Huo\altaffilmark{1}, 
        Sui-Jian Xue\altaffilmark{1}, 
        Haiguang Xu\altaffilmark{2},
        Gordon Squires\altaffilmark{3},
        Piero Rosati\altaffilmark{4} }
\altaffiltext{1}{National Astronomical Observatories, 
Chinese Academy of Sciences, Beijing 100012, China; 
Email: hzy@bao.ac.cn}
\altaffiltext{2}{Department of Physics, Shanghai Jiao Tong University, 
1954 Huashan Road, Shanghai 200030, China}
\altaffiltext{3}{SIRTF Science Center, California Institute of Technology, 
Pasadena, CA 91125 }
\altaffiltext{4}{European Southern Observatory, Karl-Scharzschild-Strasse 2, 
D-85748 Garching, Germany}

\begin{abstract}

We present joint X-ray and optical observations of the high redshift 
($z\simeq0.83$) lensing cluster CLJ~0152.7$-$1357 made
with the {\itshape Chandra\/} X-ray Observatory and the {\itshape Keck} 
telescope. We confirm the existence of significant substructure at 
both X-ray and optical wavelengths in the form of two distinct clumps, 
whose temperatures are 
$6.6_{-1.5}^{+2.4}$ keV 
and 
$5.7_{-1.6}^{+2.9}$ keV,
respectively. The X-ray surface brightness profiles of the two 
clumps can be fitted by either a single $\beta$-model or an NFW-like 
profile; the latter giving better fits to the central regions. We 
find that the X-ray derived mass of this cluster is in good agreement 
with independent lensing measurements. While its appearance indicates 
that the cluster has not reached a dynamical equilibrium state, its X-ray 
luminosity $L_{X}$, temperature $T$ and dynamical mass $M$ are consistent 
with the well-defined $L_{X}$-$T$ and $M$-$T$ relations for low-redshift 
galaxy clusters, which suggests that the dynamical properties of the clusters 
have remained almost unchanged since $z\approx0.8$. 
\end{abstract}

\keywords{cosmology: observations -- galaxies: clusters: general -- galaxies:
clusters: individual ( CLJ~0152.7$-$1357 ) -- X-ray: galaxies: clusters -- 
gravitational lensing}

\section{Introduction}
Clusters of galaxies are the largest gravitationally bound systems in
the universe, which formed recently in the hierarchical clustering
scenario. Among them, massive clusters at high redshifts are of
particular interest, being the most sensitive probes of the formation
and evolution of cosmic structures. Robust constraints on models of
structure formation and underlying cosmological parameters can be
obtained through accurate X-ray measurements of the density,
temperature, and metallicity of the intra-cluster medium (ICM)
(e.g. Bahcall \& Cen 1992; Oukbir \& Blanchard 1997; Haiman, Mohr \&
Holder 2000), as well as from studies of their scaling relation,
e.g. involving X-ray temperature, luminosity and total mass.  In
distant clusters where lensing effects are observed, one has the
ability of comparing the mass distribution derived by the X-ray data
with that inferred by the lensing analysis (e.g. Tyson, Valdes \& Wenk
1990; Kaiser \& Squires 1993), which is independent of the dynamical
equilibrium assumption, thus testing systematics inherent in both
methodologies.  CLJ~0152.7$-$1357 is an irregular, lensing cluster
with apparent substructures. The X-ray emission from its hot ICM was
first detected with {\itshape Einstein\/} IPC in 1980. Due to its
complex morphology, the source was not formally identified as a
cluster in the {\itshape Einstein\/} Extended Medium-Sensitivity
Survey (EMSS). This same source was easily discovered as an extended
source in serendipitous cluster surveys based on {\it ROSAT}-PSPC archival
data, namely {\it ROSAT} Deep Cluster Survey (RDCS) (Rosati et al. 1998), 
Wide Angle {\it ROSAT} Pointed Survey (WARPS) (Ebeling et al. 2000),
Serendipitous High-Redshift Archival {\it ROSAT} Cluster (SHARC) 
(Romer et al. 2000), and was optically identified as one of the
most rich and X-ray luminous distant clusters at
$z>0.8$. CLJ~0152.7$-$1357 was observed with {\itshape BeppoSAX\/} in
1998 (Della Ceca et al. 2000; hereafter D00). The authors reported an
average gas temperature of $kT=6.46_{-1.19}^{+1.74}$ keV and a
metallicity of $A=0.53_{-0.24}^{+0.29}$ of the solar value. However,
these results need to be confirmed because the {\itshape BeppoSAX\/}
spectrum may have been polluted by background/foreground sources that
cannot be distinguished from the diffuse X-rays of the cluster itself
due to the relatively poor spatial resolution provided by
{\itshape BeppoSAX\/}.

Using {\itshape Chandra\/} data, Maughan et al. (2003) resolved 
CLJ~0152.7$-$1357 into two massive subclusters and found their temperatures 
to be $5.5^{+0.9}_{-0.8}$ keV and $5.2^{+1.1}_{-0.9}$ keV respectively. They 
suggested that the cluster was in process of merging. In this paper, we 
present a joint study combining the same {\itshape Chandra\/} data set with
deep optical imaging data obtained with the {\itshape Keck\/} telescope. We 
evaluate and compare independent X-ray and lensing mass estimates. In \S2,
we describe the observations and data reduction. In \S3, we present analysis 
of the X-ray and optical data respectively, investigate and compare the 
different methods of mass determination. In \S5 and \S6, we summarize and 
discuss these results.

Throughout the paper, we adopt $H_0$ = 65 km s$^{-1}$ Mpc$^{-1}$, 
$\Omega_{\rm m}=0.35$ and $\Omega_{\Lambda} =0.65$. Thus, $1\arcmin$ at the 
distance of the source corresponds to 479.8 kpc. All uncertainties are 
quoted at the 90\% confidence level unless otherwise mentioned.

\section{Observations and Data Reductions}
\subsection{X-Ray Observations with the {\it Chandra} Observatory}

CLJ~0152.7$-$1357 was observed with {\itshape Chandra\/} ACIS
instrument for 36.5 ks on September 8, 2001. CCD chips I0-3 and S2-3
were operated in timed exposure mode and the data stream was
telemetered in faint mode. Diffuse X-ray emission with two X-ray
clumps, spatially coincident with two galaxy overdensities in the
optical image, were clearly detected on the I3 chip. The spectra
and images analyzed in this work were all obtained with this chip.

We followed the standard procedure by using the CIAO 2.3 package to reduce
the data. We excluded those data acquired during a short period (0.4 ks) 
when the occurrence of flares raised the background count rate to more than 
1.2 times the mean value. We cleaned the event file for bad pixels and then 
filtered it with the standard {\it ASCA} grades (G 02346). This left about
36.1 ks of screened data for imaging and spectral analysis.

Twelve point-like sources have been detected using the tool {\itshape
wavdetect\/} packaged in CIAO 2.3 and were excluded from the analysis 
of the diffuse X-ray emission. A comparison of X-ray and optical
positions of 5 point-like sources embedded in the X-ray halo of the
cluster show that the astrometric accuracy is better than 1.$\arcsec$0, 
with no need of registration. We estimate that the background- and point 
source-subtracted 0.3--10 keV count rate of the cluster is $0.077\pm0.006$ 
cts s$^{-1}$ with a $4\arcmin$ aperture. In the same band, the contribution 
of the embedded point-like sources is $0.023\pm0.001$ cts s$^{-1}$.

\subsection{Optical Observations with the {\it Keck} Telescope}
CLJ~0152.7$-$1357 was observed with the {\itshape Keck} telescope as 
a part of a program to image high-$z$ galaxy clusters drawn from the 
RDCS (Rosati \etal 1995, 1998). The 
operating detector was the Low Resolution Imaging Spectrograph 
(LRIS -- Oke \etal 1995) at Cassegrain focus, with a Tektronix 
$2048\times2048$ CCD ($0\farcs215$ per pixel).

The strategy of the observations was to obtain $\simeq 3$~hour integrations in
the R-band in sub-arcsecond seeing conditions. The depth of the R-band
observations was set by the goal of obtaining a sufficiently high
surface density of faint, distant galaxies to allow a significant
detection of weak gravitational lensing shear by a massive cluster. 
When the seeing conditions deteriorated, or the observations were 
affected by the Moon, we switched to B-/V- and I-band observations 
respectively. A summary of the observations is given in Table 1.

The image processing and lensing analysis followed the standard techniques
described below. Two-amp mode was used and a second order Legendre
polynomial was fit to the overscan region for each amplifier on an
image-by-image basis. A median bias was created from each night's
observations (comprising of 10 exposures) and subtracted from each
overscan corrected image. We created a median superflat by using all
of our science observations on each night to correct for pixel-to-pixel
variations in the response of the CCD.

Each individual exposure was corrected for optical distortion before
creating the final, coadded image as follows. A set of stars common to
our images and those of Digital Sky Survey was selected, and a second 
order spatial transformation applied to bring our images into an 
orthographically projected astrometric frame. The final median image was 
then created by stacking the transformed images.

Our object detection and analysis technique is based on that developed
by Kaiser, Squires \& Broadhurst (1995; hereafter KSB). For gravitational 
lensing analyses, there have been several investigations to determine the 
optimal method for measuring galaxy shapes, correcting for non-gravitational 
shape distortions, and calibrating gravitational shear estimates from 
galaxy shape data (e.g., KSB with extensions developed by Hoekstra \etal 
1998; Luppino \& Kaiser 1997; Kuijken 1999; Kaiser 2000). For ground based 
data, and the relatively small PSF encountered here, the standard KSB + 
Luppino/Kaiser (LK) algorithm is adequate for correcting point spread 
function anisotropies and for the losses due to seeing and pixelization.

\section{Analysis of the {\it Chandra} Data}
\subsection{X-ray Morphology and Point-like Sources}
Both the previous X-ray studies and our optical observations as reported
below indicate that CLJ~0152.7$-$1357 has a complex morphology exhibiting 
two major concentrations. This is clearly visible in {\it Chandra\/}
broadband (0.3-10 keV) image (Fig. 1). This image has been corrected for 
exposure and smoothed using the CIAO tool {\it csmooth\/} with a minimum 
significance of 3 and a maximum significance of 5. The diffuse X-ray 
emissions cover a $1'.9 \times 1'.2$ ($0.9 \times0.6$ Mpc$^2$) region, 
with two clear peaks at 
RA = 01$^h$52$^m$44$^s$.1, 
Dec = $-$13$^{\circ}$57$^{\prime}$22$^{\prime\prime}$.3 
(J2000; northeastern clump), 
and 
RA = 01$^h$52$^m$39$^s$.7, 
Dec =$-$13$^{\circ}$58$^{\prime}$30$^{\prime\prime}$.1 (J2000; southwestern
clump), respectively. The projected distance between the peaks is about
$1.5^{\prime}$ (0.7 Mpc). Both clumps are reasonably symmetrically shaped
and only slightly elongated in the northeast-southwest direction. This
suggests that the cluster is in a pre-merger state. The NE clump is larger 
in extension but less luminous than the SW clump.

In Figure 2, we show the R-band image obtained with the {\it Keck\/} 
telescope in a 4 ks exposure, on which the {\it Chandra\/} X-ray intensity 
contours are overlaid in logarithmic scales, and blow-up of the two cores in 
the {\it Keck} image, possible strong lensing features around the NE and SW 
clumps are visible in the deep {\it Keck} images. In the R band, the spatial 
distribution of galaxies also shows two similar concentrations.  However, we 
notice that the optical peaks do not exactly coincide with the X-ray peaks. 
The offset are $6.0\arcsec$ and $4.9\arcsec$ for the NE and SW clumps,
respectively. This further supports the conjecture that the cluster is
not in a dynamical equilibrium state, and may be undergoing a merger.

A close inspection of the {\it Chandra\/} and Keck images show a third
X-ray faint clump to the East of the major structure, just north of a
bright star. An overdensity of galaxies as red as those in the two
main clump is apparent in the {\it Keck} color images.

Optical counterparts of the 5 embedded X-ray point-like sources have
also been identified (Table 2 and Fig. 2). These sources are too faint
to allow a study of their X-ray spectra, we therefore calculate their
hardness ratios defined as $H = {C_{\rm H} - C_{\rm S} \over
  C_{\rm H} + C_{\rm S}}$, where $C_{\rm S}$ and $C_{\rm H}$ are the
net counts in 0.3--1.5 keV and 1.5--7 keV,
respectively. We find that the obtained hardness ratios differ
greatly, inferring that these sources are of different
origins. Besides their optical positions, no further information about
the optical or radio counterparts of these point-like sources was
found in the literature or online databases.

\subsection{Spectral Fits}
We extracted the 0.5-8 keV spectra and analyzed them with XSPEC (v11.2.0,
Arnaud 1996) 
software. Background spectra were extracted from source-free regions 
adjacent to the cluster on the ACIS I3 chip. Corrections for the 
degradation of the quantum efficiency at low energies were made. 
We also computed hardness ratios for the entire 
cluster and the two clumps separately. The hardness ratios were defined in 
the same way as in \S3.1 except that the soft and hard bands were this time 
chosen to be 0.5--2 keV and 2--8 keV, respectively. We find that the hardness 
ratios of the two clumps are consistent with each other, suggesting 
a similar temperature (Table 3).

{\bf The Entire Cluster:} We extracted the global spectrum of the
entire cluster by combining two circular regions, one centered at the
peak of each clump. We fitted the combined spectrum with the absorbed
MEKAL model designed for photoelectric absorbed emission emitted from
isothermal gas in collisional equilibrium. In the spectral fits, we
fixed the column density of the neutral hydrogen to the Galactic value
($1.55 \times 10^{20}$ cm$^{-2}$; Dickey \& Lockman 1990). The
observed spectrum and the best-fit model are showed in Figures 3--4
and Table 4, where results from previous work are also listed for
comparison. The best-fit gas temperature is $6.5_{-1.3}^{+1.7}$ keV,
or $6.2^{+1.6}_{-1.1}$ keV if the ICM abundance is fixed at 0.3 of the
solar value. With our best-fit model, we estimate that the unabsorbed
flux in 0.5--2 keV is $f_{0.5-2 \rm keV} = 1.27_{-0.11}^{+0.10}\times
10^{-13}$ erg cm$^{-2}$ s$^{-1}$, corresponding to an isotropic 0.5--2
keV luminosity of $L_{\rm X} = 3.50^{+0.29}_{-0.30} \times10^{44}$ erg
s$^{-1}$. In the 2--10 keV band, the unabsorbed flux and luminosity
are $f_{2-10 \rm keV} = 1.43_{-0.13}^{+0.12} \times 10^{-13}$ erg
cm$^{-2}$ s$^{-1}$ and $L_{\rm X} = 6.30_{-0.56}^{+0.52}
\times10^{44}$ erg s$^{-1}$ respectively.

The measured {\it Chandra} temperature is consistent with the {\it ROSAT} 
and {\it BeppoSAX} measurements (D00), while the calculated X-ray fluxes 
are lower by 42\%\/ and 38\%\/ than those obtained with {\itshape ROSAT\/} 
($f_{0.5-2 \rm keV} = 2.2 \pm 0.2 \times 10^{-13}$ erg cm$^{-2}$ s$^{-1}$)
and {\itshape BeppoSAX} ($f_{2-10 \rm keV} = 2.3 \pm 0.5 \times 10^{-13}$ 
erg cm$^{-2}$ s$^{-1}$) respectively. We found that the difference is 
primarily caused by contamination from embedded point-like sources which 
were not subtracted in the previous analyses. 
We then estimate the contribution 
of point-like sources by fitting their global spectrum with the same model 
as cluster (in order to compare with the results from {\it ROSAT} and
{\it BeppoSAX}), and find their unabsorbed flux is $f_{0.5-2 \rm keV} = 
6.6^{+0.7}_{-0.7} \times 10^{-14}$ erg cm$^{-2}$ s$^{-1}$ and
$f_{2-10 \rm keV} = 9.2^{+0.7}_{-0.7} \times 10^{-14}$ erg cm$^{-2}$ 
s$^{-1}$. Therefore, the sum of flux from cluster and point-like sources from
the {\it Chandra} data is consistent with the {\it ROSAT} and {\it BeppoSAX}
measurements. 

{\bf The Two Clumps:} We extracted the spectra of the two clumps
separately from circular regions centered on the peak of each
clump. Again, the background spectra were extracted from adjacent
source-free regions, and the spectra were fitted with the absorbed
MEKAL model.  Spectra and best fit models are shown in
Figure 5. The best-fit temperatures, obtained by fixing the $N_H$
absorption to the galactic value and the metal abundance to the
canonical value of 0.3 solar, are $6.3_{-1.3}^{+2.2}$ keV for
the NE clump, and $5.7_{-1.6}^{+2.8}$ keV for the SW clump (Table
5). If the abundance is left free, the gas temperatures do not change
significantly. At the 90\%\/ confidence level, the temperatures of the
two clumps are consistent with each other. The unabsorbed 2--10 keV
flux of the NE and SW clumps are $8.8_{-1.0}^{+0.9} \times 10^{-14}$
erg cm$^{-2}$ s$^{-1}$ (61$\pm$7\% of the entire cluster) and
$5.2^{+0.9}_{-0.8} \times 10^{-14}$ erg cm$^{-2}$ s$^{-1}$,
respectively.

\subsection{Radial Surface Brightness Profiles}
Using the 0.3--10 keV image corrected with a weighted exposure map, we
extracted the azimuthally-averaged radial surface brightness profiles 
of the two clumps separately. Embedded
point-like sources were masked out and the background was also
fitted in this procedure.

{\bf $\beta$-Model:} First we attempt to describe the surface brightness
profiles with a single $\beta$-model, which takes the form
\begin{equation}
S(r)=S_0\left[1+(\frac{r}{r_{\rm c}})^2\right]^{-3\beta+0.5}+S_{\rm B},
\end{equation}
where $r_{\rm c}$ is the core radius, $\beta$ is the index parameter,
$S_{\rm B}$ is the background, and $S_0$ is the normalization. The
best-fit models together with the observed profiles are shown in Figure
6 and Table 6. We find that the surface brightness profiles of the outer
regions of the two clumps can be well fitted with the $\beta$-model. In
the central ($<30$ kpc) regions, however, the observed profiles are
underestimated. The calculated reduced $\chi^2$ is 1.36 for the NE clump 
and 1.22 for the SW clump. Also, we find that the NE clump has a larger 
core radius than the SW clump.

We have tried to improve the fit of the central regions of the two
clumps, where a surface brightness excess is visible, by adding a
second $\beta$-model. Such a method has been successfully applied for
example in ``cooling flow'' clusters. The poor statistics in our case,
however, prevent us from constraining the additional parameters of the
double $\beta$-model.

{\bf NFW-like Model:} An alternative way to compensate for the central 
brightness excess is to utilize models that are more centrally concentrated 
than the $\beta$-model. To this aim, we have fitted the observed 
surface brightness profiles with the model expected for the gas distribution 
following an NFW-like halo profile as suggested in cosmological N-body 
simulations (Navarro, Frenk \& White 1995). In this case, 
the gas distribution has an 
analytic form of (Makino, Sasaki \& Suto 1998; Wu 2000)
\begin{equation} 
\tilde{n}_{\rm gas}(x)=\frac{(1+x)^{\alpha/x}-1}{e^{\alpha}-1},
\end{equation}
which, in order to avoid the divergence in the computation of X-ray surface 
brightness, has been normalized so that 
$\tilde{n}_{\rm gas}(x) \equiv [n_{\rm gas}(x)-n_{\rm gas(\infty)}]/[n_{\rm 
gas}(0)-n_{\rm gas}(\infty)]$ where $x=R/r_{\rm s}$, $\alpha=4 \pi G \mu 
m_{\rm p} \rho_{\rm s} r^2_{\rm s}/kT$, $\mu=0.59$ is the average molecular 
weight, and $\rho_{\rm s}$ and $r_{\rm s}$ are the characteristic density 
and length, respectively. The surface brightness profile can correspondingly 
be calculated in term of $S(x) \propto \int^{\infty}_{x}\tilde{n}^2_{\rm 
gas}(x)\,{\rm d}l$, where the integral is performed along the line of sight. 
This yields
\begin{equation}
S(x)=-S_0 \int^{\infty}_{x_0} \frac{\sqrt{x^2+x_0^2}}{x}
\left[(1+x)^{\alpha/x}-1\right]
(1+x)^{\alpha/x}\left[\frac{1}{1+x}-\frac{{\rm ln}(1+x)}{x}\right] 
{\rm d}x+ S_{\rm B},
\end{equation}
where $x_0=r/r_{\rm s}$ is the dimensionless radius.

The best-fit surface brightness profiles predicted by the NFW-like model are
presented in Figure 7 and Table 6. The calculated reduced $\chi^2$ is 1.59 
for the NE clump and 1.46 for the SW clump. It appears that the NFW-like 
model does not give a better global fit to the surface brightness profiles 
of the two clumps; however, it does improve the fit to the central regions.

\subsection{X-Ray Mass Determinations}
With the information obtained with the X-ray data, we have estimated the
gravitational mass of CLJ~0152.7$-$1357 in three ways. First, we calculate 
the virial mass of the cluster by using the cosmic virial theorem (Bryan \& 
Norman 1998)
\begin{equation}
kT=\frac{GM^{2/3}\mu m_{\rm p}}{2\beta}\,
   \left[\frac{H^2(z)\Delta_{\rm c}}{2G}\right]^{1/3}
  =1.39\,f_{\rm T}\,(\frac{M}{10^{15}
   M_{\odot}})^{2/3}\,(h^2\Delta_{\rm c}
   E^2)^{1/3}\,{\rm keV},
\end{equation}
where the normalization factor is $f_{\rm T}=0.8$, the overdensity
parameter is $\Delta_{\rm c}=200$, and 
$E(z) = h(1+z) \{\Omega_{\rm M}z+1+\Omega_{\Lambda}[(1+z)^{-2}-1]\}^{1/2}$. 
Because the {\itshape Chandra\/} temperature of the whole cluster is 
$6.5^{+1.7}_{-1.3}$ keV, the calculated virial mass is 
$9.18^{+3.82}_{-2.61}\times10^{14} M_{\odot}$.

Second, we calculate the mass distribution by assuming that the cluster
is in hydrostatic equilibrium and the gas is isothermal. In this case
the total mass within a radius $R$ can be expressed as
\begin{equation}
M_{\beta}(R)=-\frac{kTR^2}{G \mu m_{\rm p}n(R)}\frac{{\rm d}n(R)}{{\rm d}R},
\end{equation}
where $n(R)$ is the electron number density that can be inferred from
the observed surface brightness profile with the best-fit $\beta$-model. 
The calculated radial mass distribution for the two clumps are shown 
in Figure 8.

The third way to determine the mass distribution is to employ the universal 
NFW profile, in this case we have
\begin{equation}
M_{\rm NFW}(R)=4 \pi \rho_{\rm s} R_{\rm s}^3
\left[{\rm ln}(1+\frac{R}{R_{\rm s}})-\frac{R}{R+R_{\rm s}}\right].
\end{equation}
The derived mass profiles of the two clumps are also shown in Figure 8. It 
turns out that the latter two methods give essentially the same mass estimates 
except for the central regions, which may be caused by the failure of the 
hydrostatic equilibrium assumption in the core of the cluster, or by 
departure from the isothermality assumption adopted in the $\beta$-model.

\section{Analysis of the Optical Data}
\subsection{Mass Determinations}
Mass determinations from weak gravitational lensing analyses are considered 
necessarily lower estimates on the true mass of the system. In contrast with
X-ray estimates, gravitational lensing offers a method for measuring 
projected cluster masses which is essentially free from 
assumptions on the dynamical state of the gravitating matter. The observed 
shear is unperturbed by the addition of constant density sheets along the line 
of sight (e.g., Gorenstein, Shapiro \& Falco 1988). Furthermore, the 
relatively small field of view of these observations necessitates a 
differential determination of the total mass. This is facilitated by the 
statistic (Fahlman \etal 1994; Kaiser, Squires \& Broadhurst 1995)
\begin{eqnarray}
\zeta(\theta_1, \theta_2)& =& 2( 1 - \theta_1^2 / \theta_2^2 )^{-1}
        \int_{\theta_1}^{\theta_2} d \ln(\theta) \langle \gamma_t
        \rangle \\ & = & \bar{\kappa}(\theta_1) - \bar{\kappa}(\theta_1
        < \theta < \theta_2) \nonumber,
\end{eqnarray}
which measures the mean dimensionless surface density, $\bar{\kappa}$,
interior to $\theta_1$ relative to the mean in an annulus $\theta_1 <
\theta < \theta_2$, and depends only on the measured galaxy shear estimates, 
$\gamma$.  To convert this to physical units (e.g. total mass), we estimated 
the angular diameter distances factored into the dimensionless surface 
density using HDF N+S photometric redshifts (Gwyn \& Hartwick 1996, Gwyn 1999).

In Figure 9, we present the radial lensing mass profile for
CLJ~0152.7$-$1357.  The central point for the profile was chosen to be
the point mid-way between the peaks of the X-ray emission (and the
main optical concentrations of the cluster galaxies). 
For comparison purposes, we compute the mass profile for a singular
isothermal sphere (SIS), $\rho \propto r^{-2}$, with the normalization set
from the local $T_X - \sigma$ relation (Holden \etal 2003). For the observed 
temperature of $T_X \simeq 6.5 \; {\rm keV}$, this yields $\sigma \simeq 1000 
\; {\rm km/s}$.  The prediction from this model is shown as the solid line 
in Figure 9, where we have corrected for the mass in the control annulus, to 
enable a fair comparison with the lensing results.
For $r \gtrsim 0.5 \Mpc$, the SIS model matches the data fairly well. However, 
at smaller radii, it tends to overpredict the observed mass profile. In an 
attempt to improve the model, 
we introduced a large core in the density profile, 
with a softening radius of $20^{\prime\prime}$, similar to the distance from 
the central point to the peaks of the X-ray emission/optical concentrations 
of cluster galaxies. The softened model matches the observations much better 
at small radii.

Mass estimates based on weak lensing techniques are differential, and
hence tend to underestimate the total mass. By making a
model-dependent assumption on the form of the total mass profile, we
can correct for the mass in the control annulus, and extrapolate the
mass to some fiducial radius, thus providing an estimate of the mass
interior to that radius.  In Figure 10, we plot the estimate for the
projected mass within $1 \;\Mpc$, where the correction and
extrapolation are calculated from the radii in Figure 9.

\subsection{Light Distribution}
To determine the cluster light distribution, we proceeded as follows: we
obtained redshifts for 15 cluster members at {\itshape Keck} in November
1998. These galaxies form a tight locus in either a $V-I$ vs. $R - I$ or
$V-I$ vs. $V - R$ color-color plot. Using the observed colors for the
spectroscopically confirmed cluster members as a guide, we identified
additional cluster galaxy candidates as those galaxies having similar
colors, yielding a catalog containing a total of 210 galaxies. The
measured R-band magnitudes were used to estimate the cluster luminosity
as
\begin{equation}
L_B = 10^{0.4 [ M_{B\odot} - R - \overline{(B-R)}_0 + DM  + K(R)]}L_{B\odot},
\end{equation}
where $M_{B\odot} = 5.48$ is the solar absolute B magnitude, $DM =
43.71$ is the distance modulus, $\overline{(B-R)}_0 = 1.81$ is the mean
color of early type galaxies at $z = 0$ and $K(R) = 1.75$ is the
K-correction (Coleman, Wu \& Weedman 1980).

We estimated the cluster zero redshift mass-to-light ratio by calculating 
$\Sigma_L$ as the mean projected luminosity density at any radius relative to 
the mean in the control annulus outside that radius (Where zero redshift
mass-to-light ratio means that color of early type galaxies at z=0 were used.).
 We then estimated $M/L=\Sigma / \Sigma_L$, which under the assumption that 
mass traces light, forms as unbiased estimate of the cluster light 
distribution (even if the cluster extends into the control annulus). The 
results are shown in Figure 11.

\section{Discussion}
\subsection{Cooling Time}
In the central regions of the clusters where the gas density is high,
the ICM will lose most of its thermal energy radiatively in a relatively
short time in the absence of effective heating sources. By using the
best-fit surface brightness profiles for the two clumps of CLJ~0152.7$-$1357, 
we estimated their cooling time with the following analytic expression
\begin{equation}
t_{\rm {cool}}=2.21 \times 10^{10} \, {\rm yr}\, (\frac{g}{1.2})^{-1}\,
(\frac{kT}{\rm {keV}})^{1/2}\,
(\frac{n_{\rm e}}{10^{-3} \rm {cm}^{-3}})^{-1},
\end{equation} 
where $g$ is the so-called Gaunt factor. The result is shown in Figure
12, where the maximum cooling regions are marked by setting the cooling
time equal to the age of the universe. The cooling radius is 150 kpc for
the northeastern clump and 168 kpc for the southwestern clump.

\subsection{Luminosity-Temperature and Mass-Temperature Relations}
We have compared our results for CLJ~0152.7$-$1357 with the
well-defined statistical scaling relations for galaxy clusters. In
Figure 13, we plot the observed $L_{\rm X}$-$T_{\rm X}$ relation for
high-redshift clusters ($z>0.7$; Holden et al. 2002) and the best-fit
$L_{\rm X}$-$T_{\rm X}$ relation for nearby clusters (Xue \& Wu
2000). Here, the X-ray luminosity is calculated by integrating within
$r_{200}$ assuming a $\beta$-model distribution. When the model
parameters $\beta$ and $r_c$ are not available, we assume $\beta=2/3$
and $r_c=0.1$ Mpc. For comparison, we also plot the {\itshape
BeppoSAX\/} result for CLJ~0152.7$-$1357 (D00).  It appears that our
measured luminosity and temperature for CLJ~0152.7$-$1357 are
consistent with the non-evolutionary scenario of the $L_{\rm X}-T_{\rm
X}$ relation, in keeping with the results of Maughan et al. (2003).

Given the mass profile of a galaxy cluster, its virial mass can be
determined by using the characteristic radius $r_{200}$, within which
the mean mass density is 200 times the current critical density of the
universe. The calculated virial mass of CLJ~0152.7$-$1357
($9.99^{+2.46}_{-1.25}\times 10^{14} M_{\odot}$) is plotted against
the X-ray temperature in Figure 14, along with the mass-temperature
relation of nearby clusters (Xu, Jin \& Wu 2001). It is apparent that
the mass and temperature of CLJ~0152.7$-$1357 obtained with the
Chandra data agree well with the relation based on the study of nearby
clusters. These findings agree with those of Maughan et al. (2003),
where the authors compared the mass and temperature of the two
subclumps with those of the local sample (Sanderson et al. 2002) and
an intermediate redshift sample (Allen, Schmidt, \& Fabian (2001).

\section{Summary}
We have analyzed {\itshape Chandra\/} and the {\it Keck\/} telescope
observations of the high redshift cluster CLJ~0152.7$-$1357
($z\simeq0.83$).  A complex substructure with two significant
concentrations is well defined in both the X-ray and optical data. The
peaks of the galaxy distribution do not coincide precisely with the
X-ray emission peaks.  This suggests that the cluster is not in
dynamical equilibrium and may be undergoing a merger event. The X-ray
temperature of the entire cluster is found to be $6.5^{+1.7}_{-1.3}$
keV, which is consistent with the results of the previous {\itshape
BeppoSAX\/} observations ($6.46^{+1.74}_{-1.19}$ keV; D00) and
Sunyaev-Zeldovich effect measurements ($8.7^{+4.1}_{-1.8}$ keV; Joy et
al. 2001).

By measuring the gas temperature and X-ray surface brightness
profiles, we have calculated the radial distributions of dynamical
mass for the two clumps in CLJ~0152.7$-$1357 using both the
$\beta$-model and the NFW-like model. Except for the central regions
of the two clumps, the two methods yield mass estimates that are
consistent at the 90\% confidence level.  In the central regions of
the two clumps, the difference between the predictions of the two
methods may be due to a departure from of the hydrostatic equilibrium
assumption, or the inaccuracy of the $\beta$-model and the
isothermality assumption.

Using deep imaging data taken with the LRIS camera at the {\itshape
Keck} telescope, we have obtained an independent mass estimate based
on a weak gravitational lensing analysis. The mass profile is consistent
with that obtained from SIS model at large readii, and consistent
with the softened model prediction at small radii. We have compared
physical properties of CLJ~0152.7$-$1357 with those of other clusters
in the literature across a large redshift range, by investigating
their X-ray mass-temperature and luminosity-temperature relations. In
both cases, we find no departure from the well-defined correlation
relations at low redshift.  Assuming that this cluster is
representative of the general galaxy cluster population, we conclude
that there is no significant evolution of the dynamical properties for
clusters out to $z\approx0.8$.

\acknowledgments

We would like to thank the Chandra X-ray Center and the NASA High-Energy
Astrophysics Science and Research Center (HEASARC).
Some of the data presented herein were obtained at the {\it W.M.
Keck} Observatory, which is operated as a scientific partnership 
among the California Institute of Technology, the University of California,
and the National Aeronautics and Space Administration (NASA). The
Observatory was made possible by generous financial support of the
W.M. Keck Foundation. This work was supported by the National
Science Foundation of China, under Grants No. 10233040 and No.
10243001, and the Ministry of Science and Technology of China, under
Grant No. NKBRSF G19990754. Support for one of us (GKS) was provided by
NASA through Hubble Fellowship Grant No. HF-01114.01-98A from the Space
Telescope Science Institute, which is operated by the Association of
Universities for Research in Astronomy, incorporated, under NASA
Contract NAS5-26555. The research described in this paper was carried
out, in part, by the Jet Propulsion Laboratory, California Institute of
Technology, and was sponsored by the National Aeronautics and Space
Administration.

\begin{figure}
\epsscale{0.8}
\figcaption{A smoothed and exposure-corrected {\itshape Chandra}
 image of CLJ~0152.7$-$1357 in 
0.3--10 keV band. Contours span the range: $3.6 \times 10^{-9}$ 
photons cm$^{-2}$ 
s$^{-1}$ pixel$^{-2}$ to $1.5 \times 10^{-8}$ photons cm$^{-2}$ s$^{-1}$ 
pixel$^{-2}$ using a logarithmic scale. The two crosses mark the two 
X-ray peaks. The field of view is $6'.56\times6'.56$, or $3.1\times3.1$ 
Mpc$^{2}$.
\label{fig-1}}
\end{figure}

\begin{figure}
\epsscale{0.9}
\figcaption{An R-band image of CLJ~0152.7$-$1357 obtained with the
 LRIS instrument at the {\itshape Keck\/} telescope with a 4 ks exposure,
 {\itshape Chandra} X-ray contours are overlaid (left panel). Blow-up of the 
two cores in the {\it Keck} image (right panel).
\label{fig-2}}
\end{figure}

\setcounter{figure}{2}
\begin{figure}
\epsscale{0.8}
\begin{center}
\includegraphics[width=10cm,angle=270]{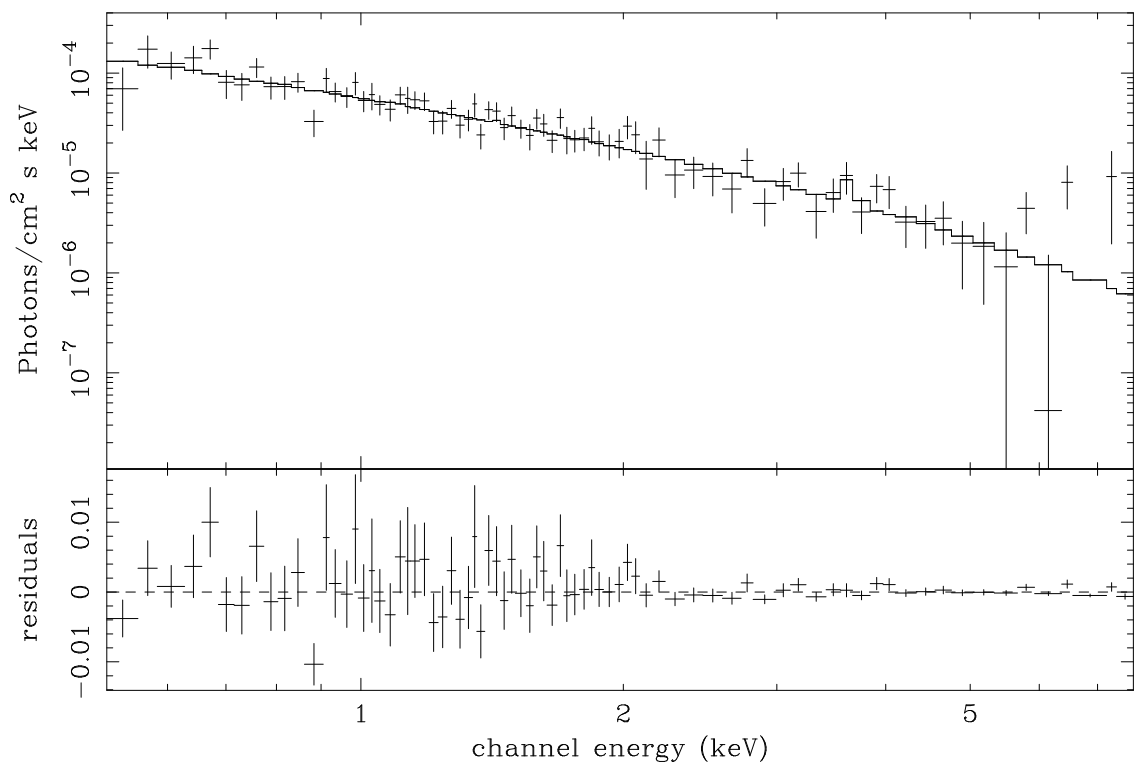}
\end{center}
\figcaption{The observed unfolded X-ray spectrum and the best-fit model for the 
entire cluster. \label{fig-3}}
\end{figure}

\begin{figure}
\epsscale{0.8}
\begin{center}
\includegraphics[width=9cm,angle=270]{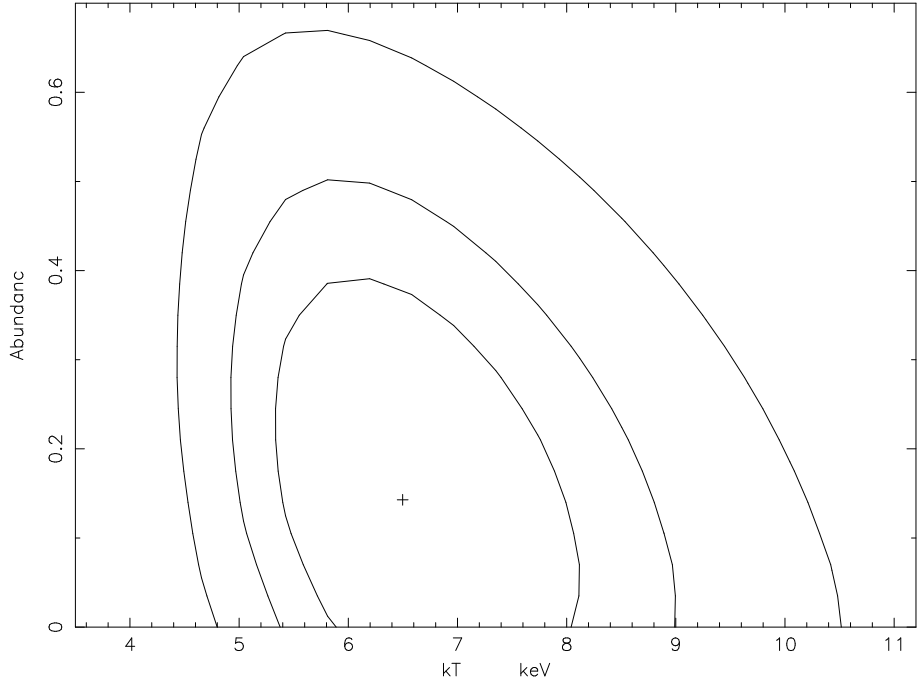}
\end{center}
\figcaption{Confidence contours (68.3\%, 90\%, and 99\%) from a spectral fit 
to the entire cluster as a function of gas temperature and metal 
abundance.
\label{fig-4}}
\end{figure}

\begin{figure}
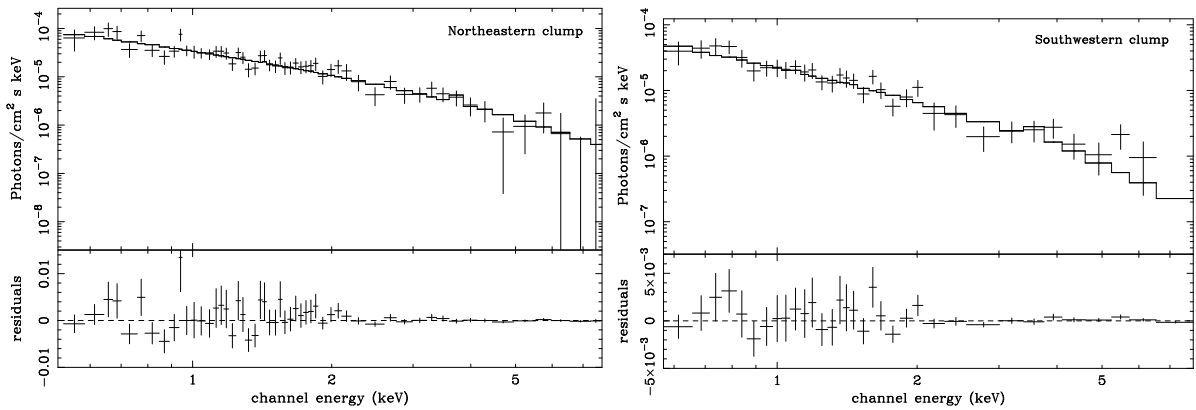

\begin{center}
\includegraphics[width=5.3cm,angle=270]{huo.fig5a.eps}
\includegraphics[width=5.3cm,angle=270]{huo.fig5b.eps}
\end{center}
\figcaption{Observed unfolded X-ray spectra and the best-fit models for the 
northeastern and southwestern clumps.
\label{fig-5}}
\end{figure}

\begin{figure}
\epsscale{1.1}
\plottwo{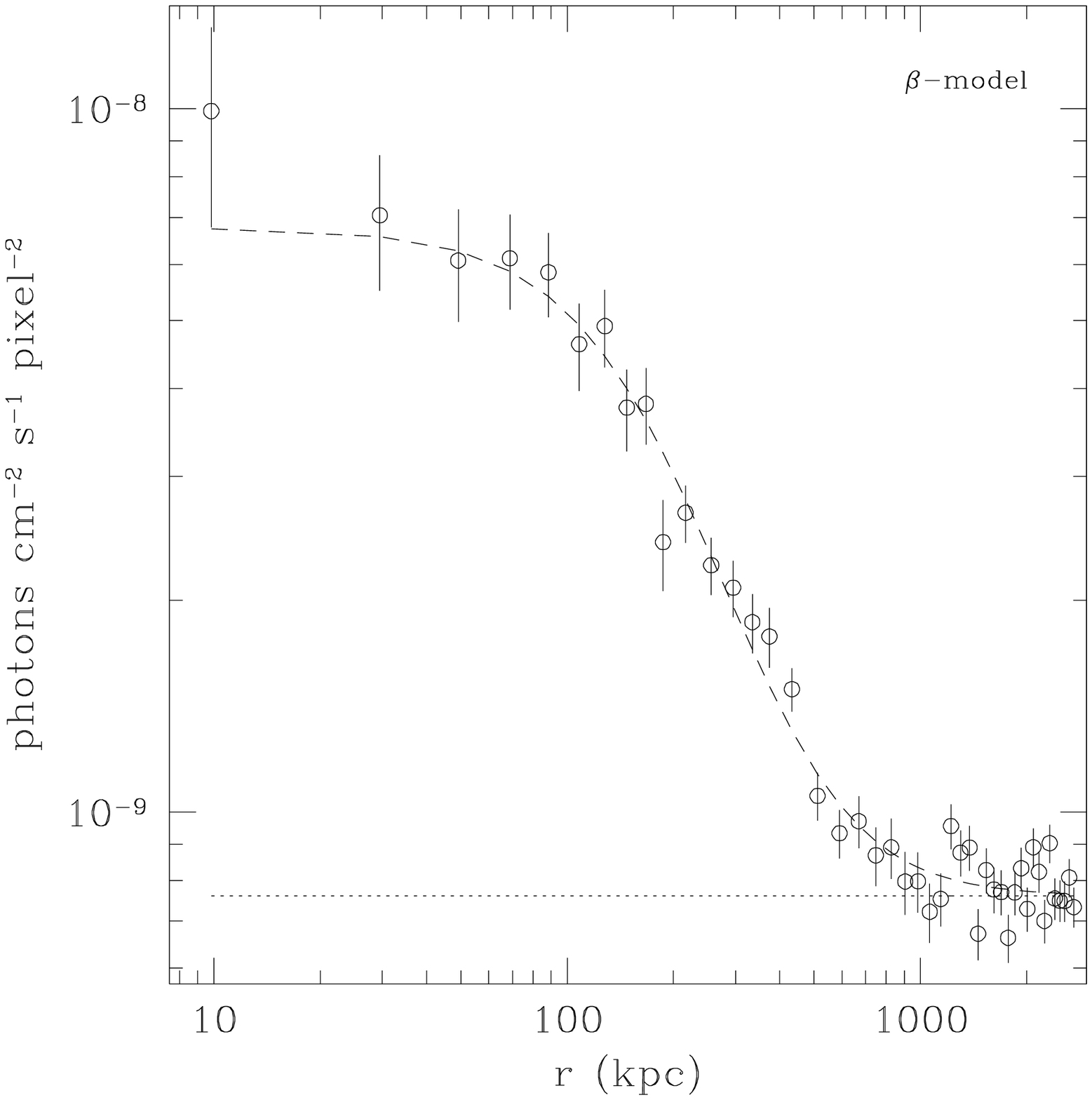}{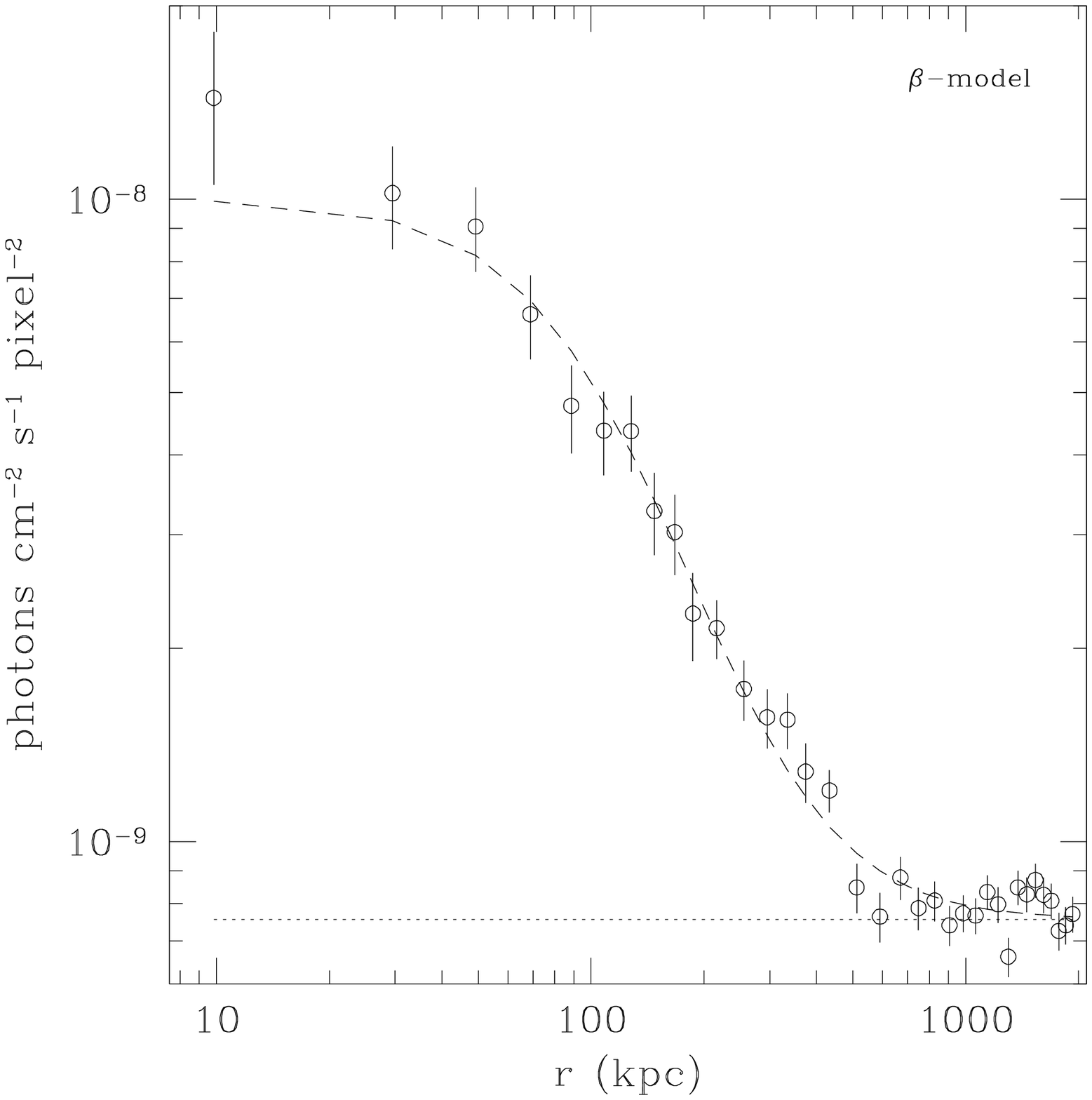}
\figcaption{Observed radial surface brightness profiles (open circle)
and the best-fit $\beta$-model (dashed line) for the northeastern clump 
(left panel) and southwestern clump (right panel). The best-fit background 
is represented by the dotted line.
\label{fig-6}}
\end{figure}

\begin{figure}
\epsscale{1.1}
\plottwo{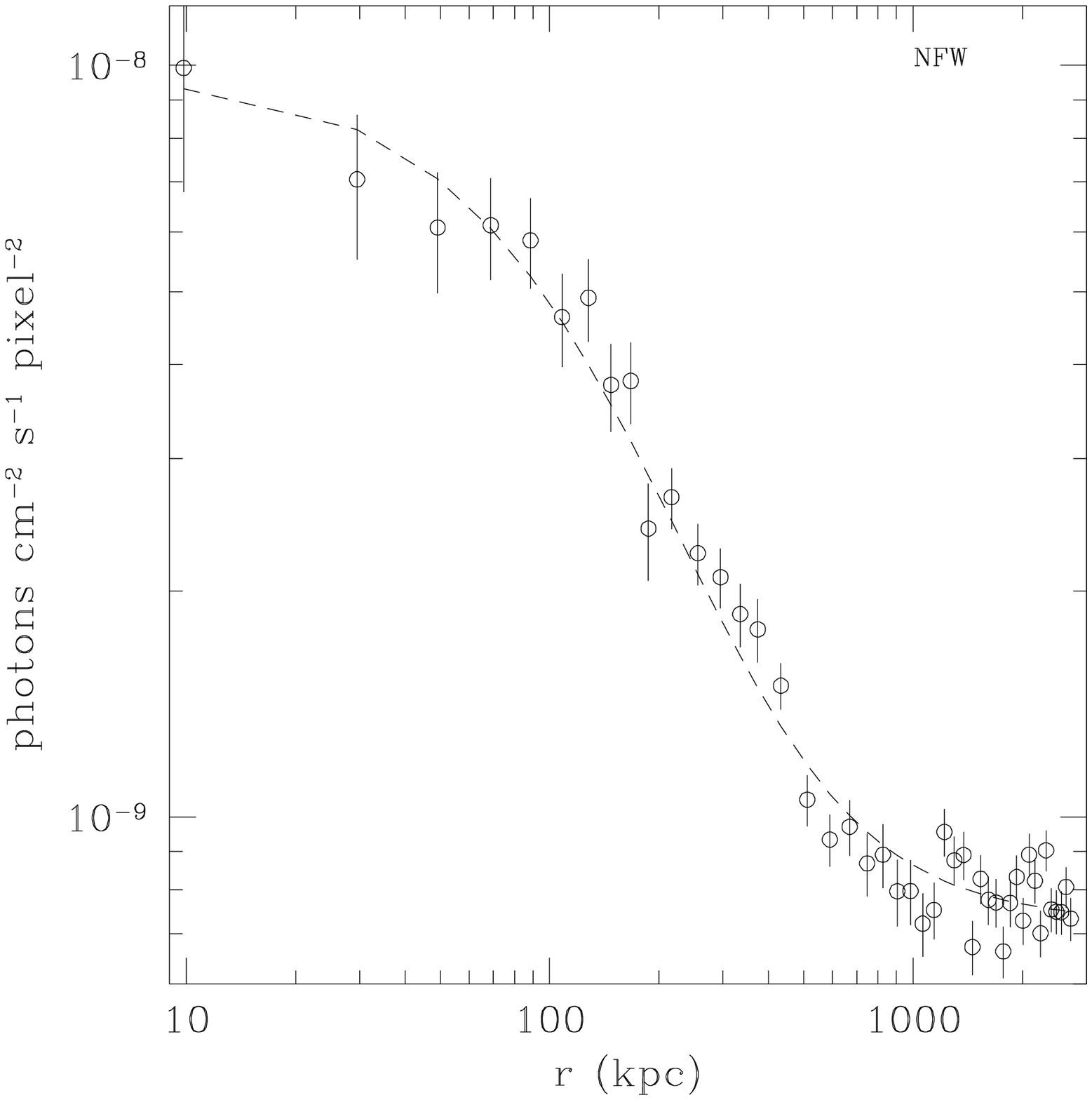}{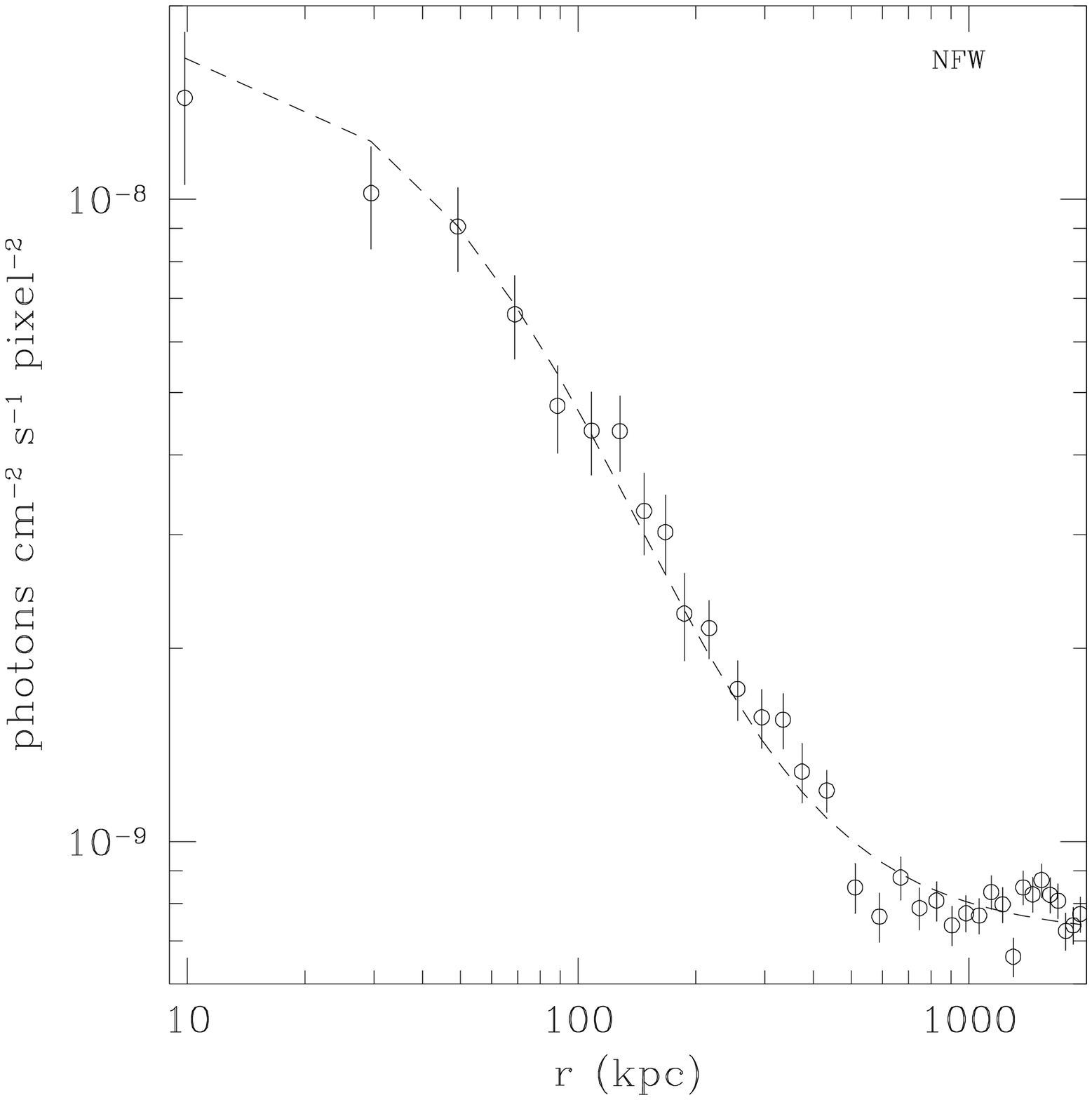}
\figcaption{Observed radial surface brightness profiles (open circle) and 
the best-fit model using an NFW-like profile (dashed line) for the northeastern clump (left panel) and southwestern panel (right panel).
\label{fig-7}}
\end{figure}

\begin{figure}
\epsscale{1.1}
\plottwo{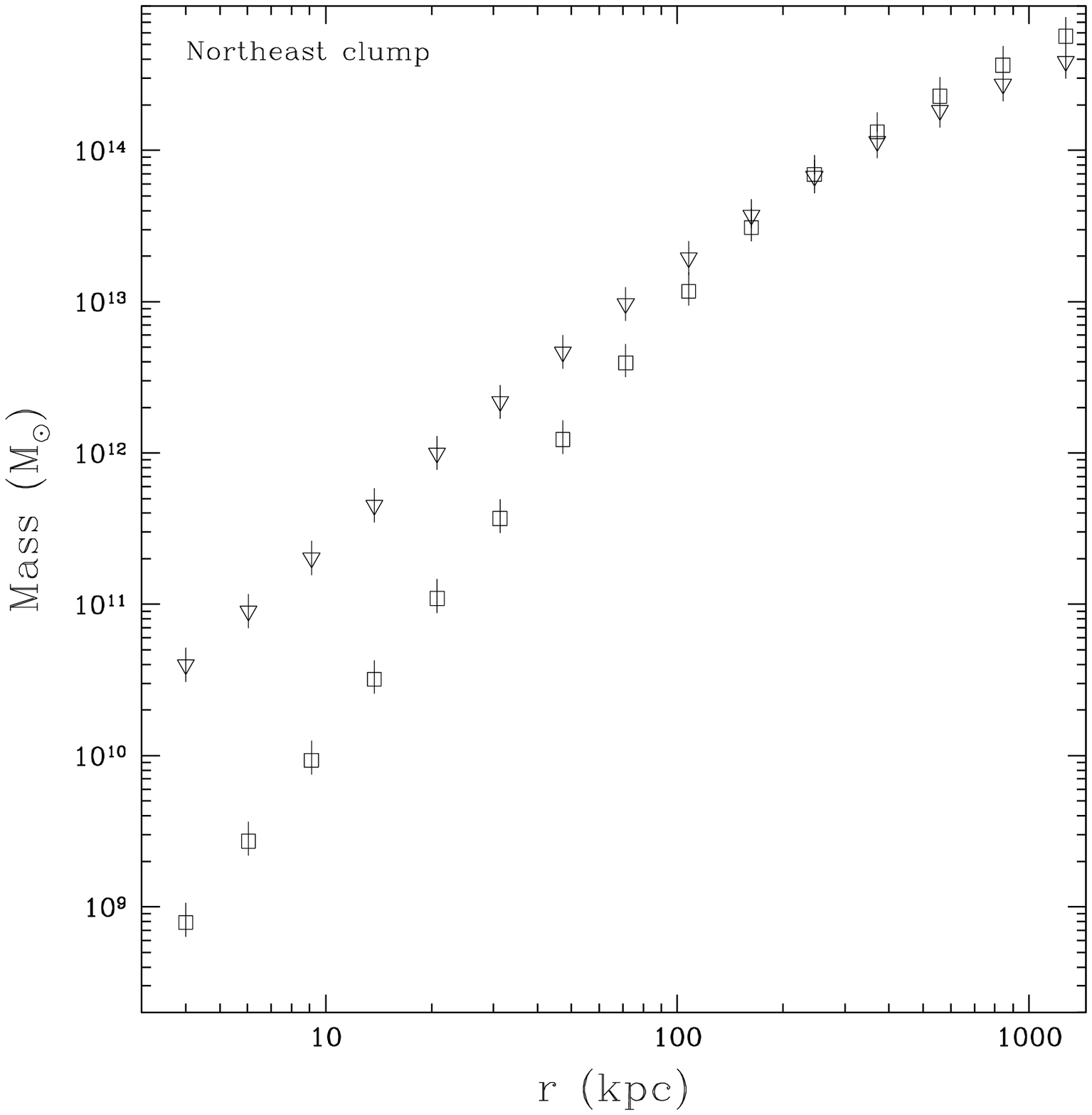}{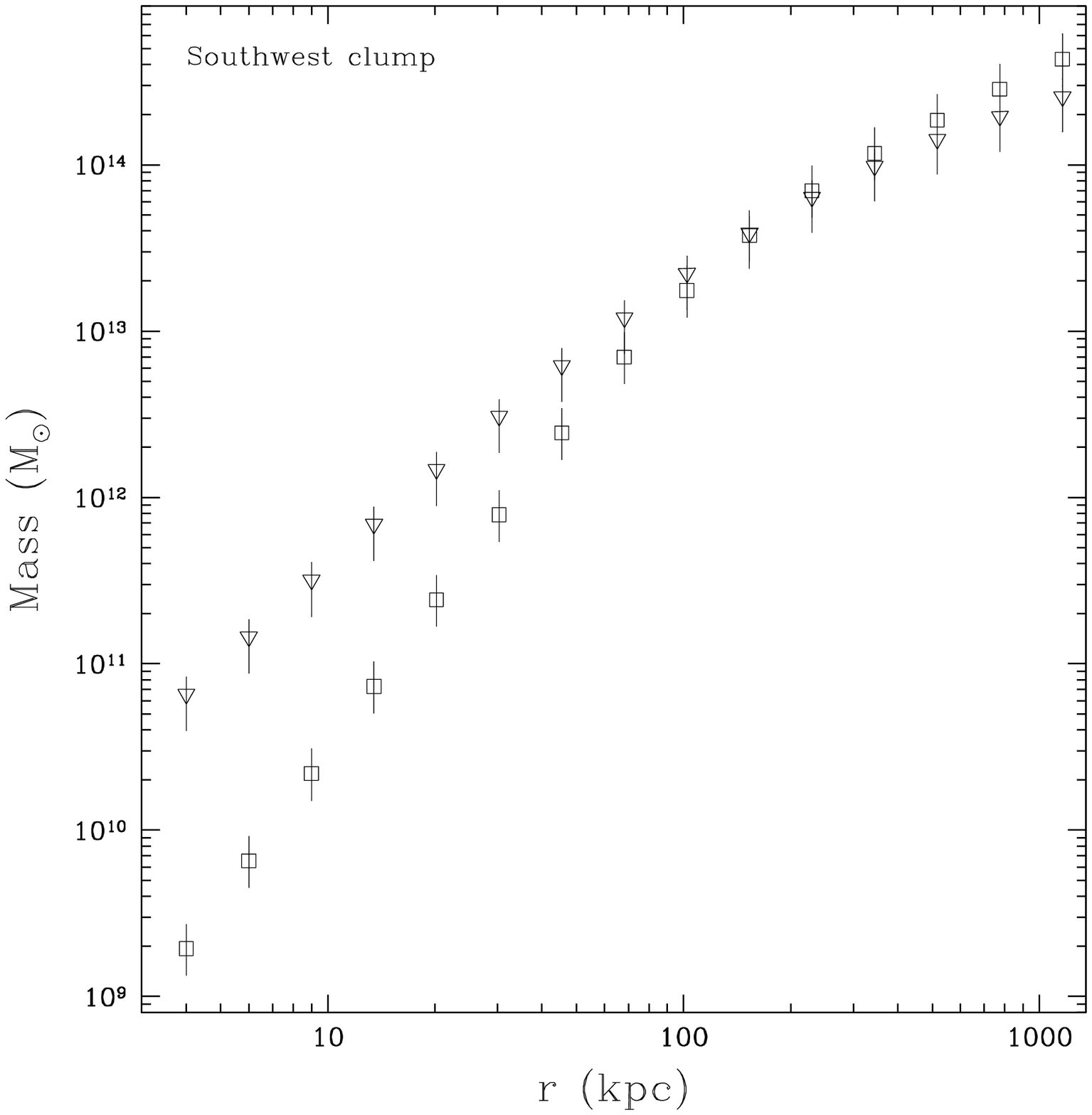}
\figcaption{Radial mass profiles of the two clumps obtained with the best-fit
$\beta$-model (square) and NFW-like profile (down triangle).
\label{fig-8}}
\end{figure}

\begin{figure}
\epsscale{0.8}
\plotone{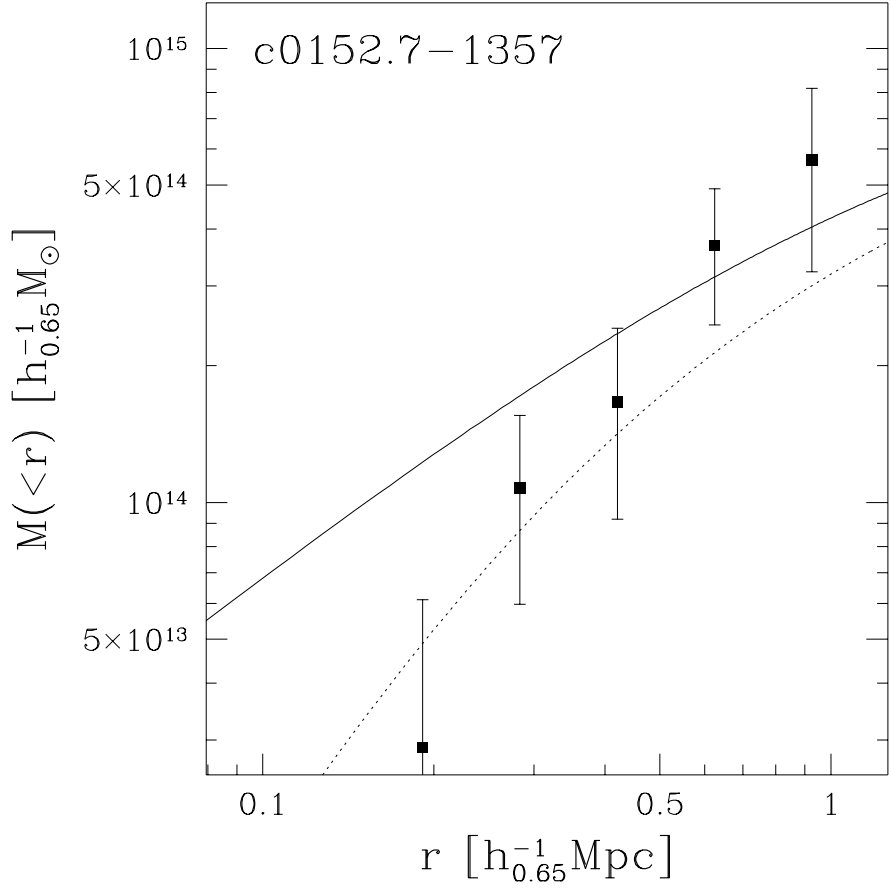}
\figcaption{Cumulative mass profile (data points) for the whole 
cluster based on the lensing inversion technique. 
The solid line represents the 
prediction for a singular isothermal sphere model, where the normalization 
has been set to the local X-ray temperature/velocity dispersion relation, 
and assumes that the mass distribution extends into the control region. 
The effect of substructure was modeled by modifying the isothermal sphere 
to include a core radius of $20^{\prime\prime}$ (dashed line), where the 
core radius was chosen to match the angular separation of the peaks of the 
X-ray emission and optical centroids of the cluster galaxy concentrations.}
\label{fig-9}
\end{figure}

\begin{figure}
\epsscale{0.8}
\plotone{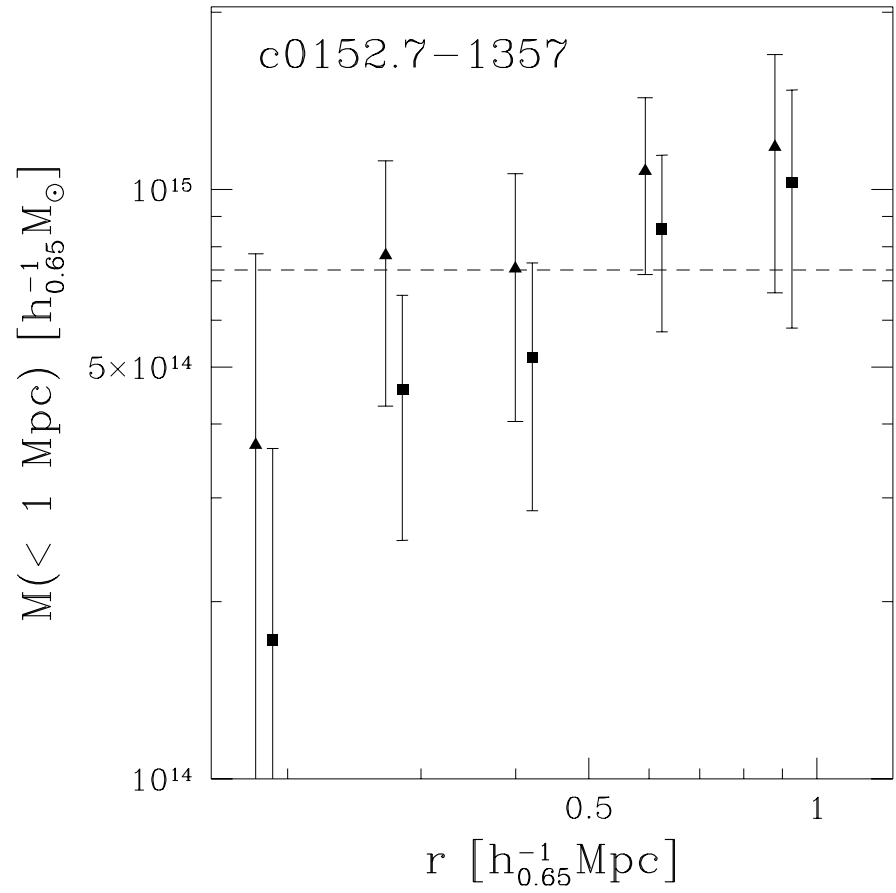}
\figcaption{The projected cluster mass within $r = 1 \, \Mpc$, 
as estimated from the weak lensing analysis at various radii. 
The dashed line is 
a prediction for a singular isothermal sphere normalized using the cluster 
X-ray temperature.}
\label{fig-10}
\end{figure}

\begin{figure}
\epsscale{0.8}
\plotone{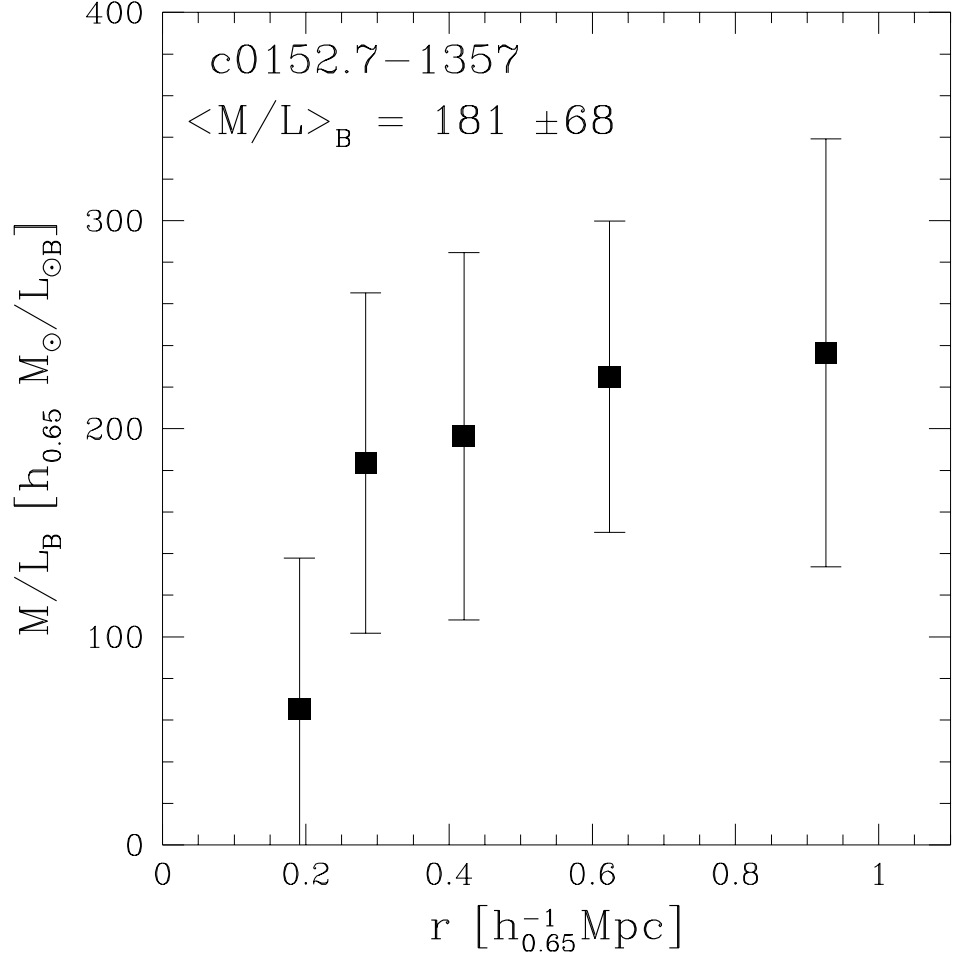}
\figcaption{The zero redshift mass-to-light ratio. The cluster light 
distribution was determined by choosing galaxies with colors similar to 15 
spectroscopically confirmed cluster galaxies ($ 0.86 < V - R < 1.75 $ and 
$ 2.40 < V - I < 3.39 $). The mass-to-light ratio was taken as the ratio of 
the radial mean surface mass density to the projected luminosity density, 
relative to the mean in the circular control region outside that radius. 
Under the assumption that mass traces light, this forms an unbiased estimate 
of the cluster light distribution (even if the cluster extends into the 
control annulus).}
\label{fig-11}
\end{figure}

\begin{figure}
\epsscale{0.8}
\plotone{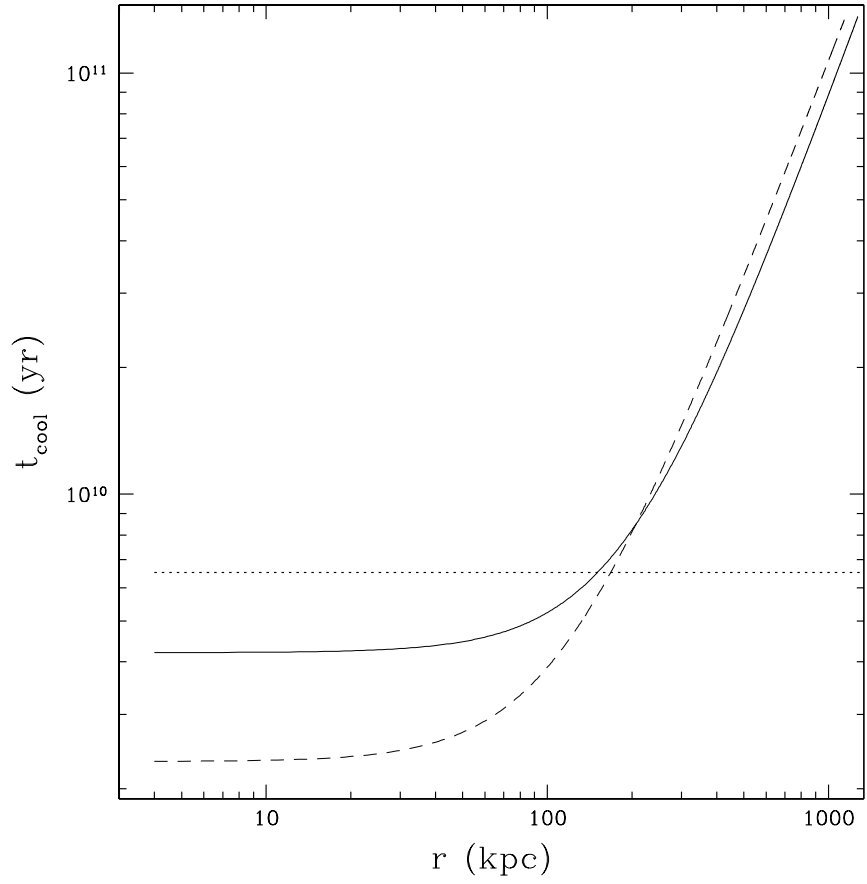}
\figcaption{Cooling time profiles for the northeastern clump (solid line) and 
the southwestern clump (dashed line). The dotted line represents the age of 
the universe at the cluster redshift.
\label{fig-12}}
\end{figure}

\begin{figure}
\epsscale{0.8}
\plotone{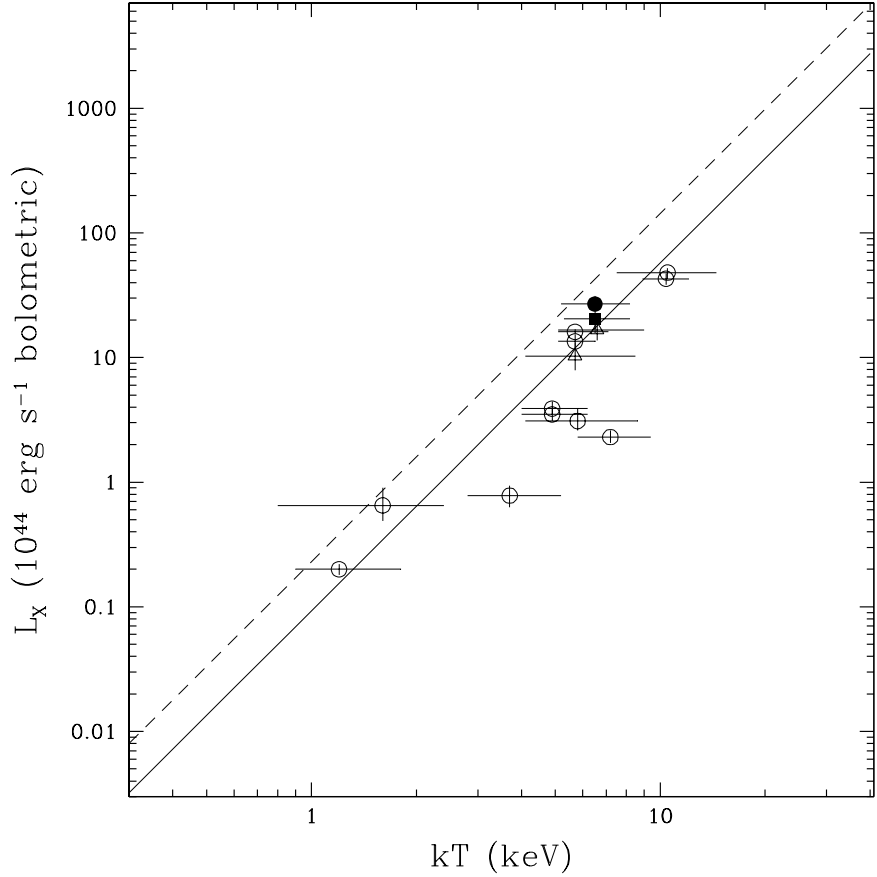}
\figcaption{Luminosity-temperature relation for galaxy clusters. Open circles: 
distant clusters and groups at $z>0.7$ (Holden et al. 2002). Solid square: 
the {\itshape BeppoSAX\/} measurements of CLJ~0152.7$-$1357 (D00). Triangles: 
the {\itshape Chandra\/} measurements of the two subclumps of 
CLJ~0152.7$-$1357. Solid circle: the {\itshape Chandra\/} measurement of the 
entire cluster. Solid line: luminosity-temperature relation for nearby 
clusters (Xue \& Wu (2000). Dashed line: the predicted evolving $L_{X}-T$ 
relation with $(1+z)^{1.5}$ scaling.
\label{fig-13}}
\end{figure}

\begin{figure}
\epsscale{0.8}
\plotone{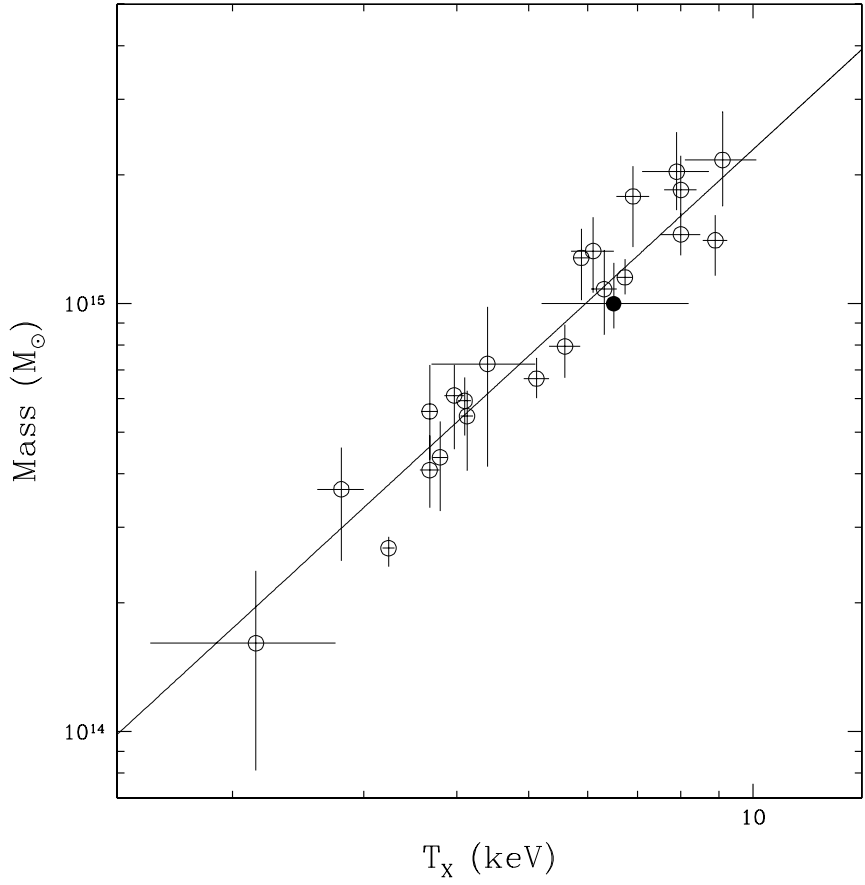}
\figcaption{Mass-temperature relation for nearby clusters (open circles, 
Xu, Jin \& Wu 2001) and our {\itshape Chandra\/} result for  CLJ~0152.7$-$1357
(solid circles).
\label{fig-14}}
\end{figure}

\clearpage

\begin{deluxetable}{ccc}
\tablewidth{300pt}
\tablecaption{Summary of the {\itshape Keck}/LRIS Observations of 
CLJ~0152.7$-$1357}
\tablehead{
\colhead{Filter} & \colhead{$t_{\rm exp}$} & {Seeing} \\
                 & \colhead{(s)}         &             }
\startdata
B & 7189   & 1\farcs18 \\
V & 1500   & 0\farcs87 \\
R & 12993  & 0\farcs80 \\
I & 1885   & 0\farcs67
\enddata
\end{deluxetable}

\clearpage
\begin{deluxetable}{cccccc}
\tablecaption{Positions, counts and hardness ratios of the 5 point-like 
sources with optical counterparts embedded in the diffuse X-ray emission}
\tablehead{
\multicolumn{1}{c}{Sources} &
\multicolumn{2}{c}{Source Position (J2000)} & \multicolumn{2}{c}{Counts} &
\multicolumn{1}{c}{Hardness Ratios}\\
\multicolumn{1}{c}{} &  \multicolumn{1}{c}{RA} & \multicolumn{1}{c}{Dec} & 
\multicolumn{1}{c}{0.3$-$1.5 keV} & \multicolumn{1}{c}{1.5$-$7 keV} &
\multicolumn{1}{c}{} }
\startdata
A & 01$^h$52$^m$45$^s$.8 & $-$13$^{\circ}$55$^{\prime}$29$^{\prime\prime}$.2 
& $11.4\pm3.5$ & $5.0\pm 2.2$ & $-0.39\pm0.06$ \nl 
B & 01$^h$52$^m$43$^s$.8 & $-$13$^{\circ}$59$^{\prime}$02$^{\prime\prime}$.2 
& $216.0\pm14.7$ & $98.3\pm10.0$ & $-0.37\pm0.01$ \nl 
C & 01$^h$52$^m$40$^s$.9 & $-$14$^{\circ}$00$^{\prime}$09$^{\prime\prime}$.1 
& $31.8\pm6.1$ & $14.7\pm4.1$ & $-0.37\pm0.04$ \nl 
D & 01$^h$52$^m$39$^s$.8 & $-$13$^{\circ}$57$^{\prime}$41$^{\prime\prime}$.5 
& $82.3\pm9.2$ & $149.0\pm12.4$ & $0.29\pm0.01$ \nl  
E & 01$^h$52$^m$34$^s$.7 & $-$13$^{\circ}$59$^{\prime}$29$^{\prime\prime}$.7 
& $23.3\pm5.1$ & $20.3\pm4.8$ & $-0.07\pm0.01$ \nl
\enddata
\end{deluxetable}

\clearpage
\begin{deluxetable}{lccc}
\tablewidth{380pt}
\tablecaption{Hardness ratios for the two clumps}
\tablehead{
\multicolumn{1}{c}{Clumps} & \multicolumn{2}{c}{Counts} &
\multicolumn{1}{c}{Hardness Ratios}\\
\multicolumn{1}{c}{} & 
\multicolumn{1}{c}{0.5$-$2 keV} & \multicolumn{1}{c}{2-8 keV} &
\multicolumn{1}{c}{} }
\startdata
Entire cluster     & $908.1\pm32.6$   & $264.5\pm23.6$ & $-0.55\pm0.02$ \nl
Northeastern clump & $551.2\pm25.3$   & $164.5\pm18.3$ & $-0.54\pm0.02$ \nl
Southwestern clump & $356.9\pm20.1$   & $100.0\pm13.8$ & $-0.56\pm0.03$ \nl
\enddata
\end{deluxetable}

\clearpage
\begin{deluxetable}{lcccc}
\tablewidth{430pt}
\tablecaption{Gas temperature and metal abundance measurements for the 
entire cluster}
\tablehead{
\colhead{Observations} & \colhead{Temperature} & \colhead{Metallicity} 
 & \colhead{ $\chi^{2}$/d.o.f} & \colhead{Ref.}\\
 & \colhead{(keV)} & \colhead{(Z$_{\odot}$)} & &  }
\startdata 
Chandra & $6.2_{-1.1}^{+1.6}$ & $0.3$(fixed) & 78.6/76 &   \nl
Chandra & $6.5_{-1.3}^{+1.7}$ & $0.14_{-0.14}^{+0.27}$ & 77.6/75 &  \nl
ROSAT   & 6 & - & - & 1 \nl
BeppoSAX & $6.46^{+1.74}_{-1.19}$ & $0.53^{+0.29}_{-0.24}$ & 16.6/21 & 1 \nl
SZ-effect & $8.7^{+4.1}_{-1.8}$ &  -  & - & 2 \nl
Chandra$^{\rm n} $ & $5.5^{+0.9}_{-0.8}$ & $0.3$(fixed) & - & 3 \nl
Chandra$^{\rm n} $ & $5.6^{+1.0}_{-0.8}$ & $0.15^{+0.22}_{-0.15}$ & 32.4/36 & 3 \nl
Chandra$^{\rm s}$ & $5.2^{+1.1}_{-0.9}$ & $0.3$(fixed) & - & 3 \nl
Chandra$^{\rm s}$ & $4.8^{+1.4}_{-0.9}$ & $0.70^{+0.72}_{-0.44}$ & 24.5/35 & 3 \nl
\enddata
\tablenotetext{n}{northeastern clump}
\tablenotetext{s}{southwestern clump}
\tablerefs{(1) Della Ceca et al. 2000; (2) Joy et al. 2001; (3) Maughan et al. 2003}
\end{deluxetable}

\clearpage
\begin{deluxetable}{cccc}
\tablewidth{370pt}
\tablecaption{Gas temperatures and metal abundances of the two clumps}
\tablehead{
\colhead{Clumps} & \colhead{Temperature} & \colhead{Metallicity}
& \colhead{$\chi^{2}$/d.o.f} \\
& \colhead{(keV)} & \colhead{(Z$_{\odot}$)} & }
\startdata
Northeastern clump & $6.3^{+2.2}_{-1.3}$ & $0.3$(fixed)           & 43.6/52 \nl
Northeastern clump & $6.6^{+2.4}_{-1.5}$ & $0.16^{+0.36}_{-0.16}$ & 43.2/51 \nl
Southwestern clump & $5.7^{+2.8}_{-1.6}$ & $0.3$(fixed)           & 28.0/32 \nl
Southwestern clump & $5.7^{+2.9}_{-1.6}$ & $0.31^{+0.55}_{-0.31}$ & 28.0/31 \nl
\enddata
\end{deluxetable}

\clearpage
\begin{deluxetable}{lccc}
\tablewidth{410pt}
\small
\tablecaption{Best-fits to the surface brightness profiles
with a $\beta$-model and an NFW-like model}
\tablehead{ 
\colhead{Model} & Parameters & \colhead{Northeastern clump} & 
\colhead{Southwestern clump}}
\startdata
& $S_0$$^{\rm a}$       & $(6.01\pm0.40) \times 10^{-9}$ &
$(9.25\pm0.72) \times10^{-9}$ \nl
& $r_c$ ( $\arcsec$ ) & $24.05\pm1.09$ & $13.90\pm0.70$ \nl
{$\beta$-model} & $\beta$  & $0.61\pm0.02$  & $0.58\pm0.02$ \nl
& $S_B$$^{\rm a}$ & $(7.61\pm0.18) \times 10^{-10}$ &
$(7.56\pm0.19) \times10^{-10}$ \nl
& $\chi^{2}$/d.o.f    & 55.68/41       & 37.67/31\nl
\tableline 
& $S_0$$^{\rm a}$ & $2.50^{+0.36}_{-0.21} \times 10^{-12}$ &
$4.90^{+0.64}_{-0.60} \times10^{-12}$ \nl
& $r_s$ ( $\arcsec$ ) & $37.50^{+4.26}_{-2.24}$ & $19.34^{+2.17}_{-1.46}$ \nl     
NFW-like model & $\alpha$ & $6.11^{+0.13}_{-0.07}$ & $6.11^{+0.10}_{-0.13}$ \nl
& $S_B$$^{\rm a}$ & $7.11^{+0.30}_{-0.31} \times 10^{-10}$ &
$6.91^{+0.35}_{-0.31} \times10^{-10}$ \nl
& $\chi^{2}$/d.o.f & 65.10/41 & 45.23/31 \nl
\enddata
\tablenotetext{a}{In units of photons cm$^{-2}$ s$^{-1}$ pixel$^{-2}$.}
\end{deluxetable}

\end{document}